# Integrating Virtual Reality and Large Language Models for Team-Based Non-Technical Skills Training and Evaluation in the Operating Room

Jacob Barker
School of Computer Science,
University of Oklahoma

Doga Demirel
School of Computer Science,
University of Oklahoma

Cullen Jackson
Department of Industrial Engineering,
Clemson University

Anna Johansson
Department of Medicine,
Beth Israel Deaconess Medical Center

Robbin Miraglia
Carl J. Shapiro Simulation and Skills Center,
Beth Israel Deaconess Medical Center

Darian Hoagland
Department of Surgery,
Beth Israel Deaconess Medical Center

Stephanie B. Jones
Department of Anesthesiology,
Northwell Health

John Mitchell
Department of Anesthesiology,
Henry Ford Health

Daniel B. Jones
Department of Surgery,
Rutgers New Jersey Medical School

Suvranu De
College of Engineering,
Florida Agricultural and Mechanical University and Florida State University

**ABSTRACT**

Although effective teamwork and communication are critical to surgical safety, structured training for non-technical skills (NTS) remains limited compared with technical simulation. The ACS/APDS Phase III Team-Based Skills Curriculum calls for scalable tools that both teach and objectively assess these competencies during laparoscopic emergencies. We introduce the Virtual Operating Room Team Experience (VORTeX), a multi-user virtual reality (VR) platform that integrates immersive team simulation with large language model (LLM) analytics to train and evaluate communication, decision-making, teamwork, and leadership. Team dialogue is analyzed using structured prompts derived from the Non-Technical Skills for Surgeons (NOTSS) framework, enabling automated classification of behaviors and generation of directed interaction graphs that quantify communication structure and hierarchy. Two laparoscopic emergency scenarios, pneumothorax and intra-abdominal bleeding, were implemented to elicit realistic stress and collaboration. Twelve surgical professionals completed pilot sessions at the 2024 SAGES conference, rating VORTeX as intuitive, immersive, and valuable for developing teamwork and communication. The LLM consistently produced interpretable communication networks reflecting expected operative hierarchies, with surgeons as central integrators, nurses as initiators, and anesthesiologists as balanced intermediaries. By integrating immersive VR with LLM-driven behavioral analytics, VORTeX provides a scalable, privacy-compliant framework for objective assessment and automated, data-informed debriefing across distributed training environments.

# 1 INTRODUCTION

Surgical teams represent complex socio-technical systems composed of interdependent professionals, surgeons, anesthesiologists, and nurses, who must coordinate seamlessly under variable clinical, temporal, and cognitive demands [1]. Effective team performance depends on continuous information exchange, shared situational awareness, and adaptive leadership, all of which are essential not only during crises but throughout routine surgical care. Non-technical skills (NTS) including communication, teamwork, decision-making, and leadership, are central to maintaining efficiency, safety, and patient outcomes across the surgical continuum.

Emergencies in the operating room (OR) amplify these demands, requiring rapid, coordinated responses under high stress and uncertainty. Elevated stress during these crises may impair performance and jeopardize patient outcomes [2], particularly when compounded by breakdowns in teamwork and communication [3]. While simulation-based training has substantially advanced technical skill acquisition, structured training for non-technical skills NTS remains limited. Furthermore, few simulators replicate the acute cognitive load and stress inherent to intraoperative emergencies. Emerging evidence suggests that incorporating realistic stressors into simulations enhances transfer of training to the intraoperative emergencies [4]. Addressing this gap is central to the evolution of surgical education, ensuring that leadership and teamwork competencies are developed alongside procedural expertise to improve patient safety and surgical outcomes.

Laparoscopic cholecystectomy (LC) is the gold standard for gallstone disease [4] and serves as one of the procedural foundations for the American College of Surgeons (ACS)/ Association of Program Directors in Surgery (APDS) Phase III Team-Based Skills Curriculum [5]. This advanced curriculum explicitly mandates simulation-based training for laparoscopic emergencies to develop crisis resource management competencies across surgical teams. The validation study by Powers et al. [6] demonstrated that structured team training for LC emergency scenarios significantly improved communication and error mitigation, yet national adoption remains limited. Bleeding occurs in approximately 10% of LC cases [7], while rare complications such as pneumothorax (0.01–0.4% incidence [8]) demand immediate, coordinated team response. The infrequent but high-stakes nature of these events, combined with the critical need for precise communication, makes LC an ideal case study for exploring how digital platforms can standardize team responses across institutions and experience levels.

Traditional facility-based simulation is the *de facto* mode for OR team training. Simulation centers employing high-fidelity physical mannequins and mock operating rooms have demonstrated efficacy in improving team communication and crisis management [9]. These programs typically require co-located teams, dedicated simulation staff, and substantial time commitments, often necessitating travel to specialized centers [10]. While effective, this model creates significant logistical barriers to widespread implementation and longitudinal skill development.

Despite established benefits, the resource-intensive nature of facility-based simulation has resulted in low national adoption of the ACS/APDS Phase III curriculum. A 2013 survey of general surgery program directors found that only 16% implemented Phase III of the curriculum, citing cost, coordinator availability, and limited personnel as primary barriers [9]. This implementation gap leaves most surgical trainees without structured opportunities to practice non-technical skills in high-fidelity crisis scenarios, limiting translation of simulation-based education into routine practice.

Virtual reality offers a geographically flexible alternative that can overcome these barriers while preserving psychological fidelity. Although VR has proven effective for individual laparoscopic psychomotor training [11], [12], its application to team-based non-technical skills remains nascent. Recent work demonstrates that multi-user VR environments can replicate the spatial and communication demands of clinical settings [13], [14], enabling distributed teams to train together without physical colocation. Consumer-grade VR systems further reduce cost barriers, making frequent, team-based practice feasible.

Existing VR training systems have largely focused on procedural skills and often depend on expensive hardware or physical simulation centers. In contrast, the Virtual Operating Room Team Experience (VORTeX) simulator leverages consumer-grade, multi-user VR to target the behavioral and cognitive dimensions of surgical teamwork. The platform's novel integration of immersive crisis simulation with a large language model (LLM)-based analytics engine enables automated, objective assessment of communication, leadership, and decision-making behaviors. To our knowledge, this is among the first applications of LLM-driven behavioral analysis to surgical team training, transforming unstructured dialogue into structured performance metrics.

The primary aim of this study is to evaluate the feasibility and perceived educational value of VORTeX for training and assessing non-technical skills within the ACS/APDS Phase III curriculum framework. The secondary aim is to validate the LLM-based analysis pipeline's capability to generate communication interaction networks that reflect expected role-dependent hierarchies during simulated emergencies. We hypothesize that (1) participants will perceive VORTeX as an effective and usable tool for enhancing teamwork under stress, and (2) LLM-generated interaction graphs will quantitatively mirror known intraoperative communication structures (surgeons as central integrators, nurses as initiators, anesthesiologists as intermediaries), supporting their use for objective assessment.

The following sections describe the design, implementation, and pilot validation of VORTeX. Section 2 details the platform architecture, professional roles, emergency scenarios, and LLM-based feedback methodology. Section 3 presents results from a pilot study conducted at the 2024 Society of American Gastrointestinal and Endoscopic Surgeons (SAGES) annual meeting, including usability metrics, participant perceptions, and analysis of LLM-generated communication graphs. Section 4 discusses implications for competency-based surgical education under the Phase III framework, and Section 5 concludes with limitations and directions for future multi-institutional implementation.

## 2 METHODS

### 2.1 System Overview

The Virtual Operating Room Team Experience (VORTeX) is a multi-user VR platform designed to train and objectively assess non-technical skills such as communication, teamwork, and decision-making in surgical teams. The platform enables participants to collaborate in shared operating room scenarios through networked VR environments, allowing geographically distributed teams to engage in realistic simulations of both routine and emergent intraoperative events.

As illustrated in Figure 1, the system comprises two core components: the **simulation system** and the **feedback engine**. The simulation engine integrates consumer-grade VR hardware, including the headset, hand controllers, and voice communication, with a network synchronization layer that ensures real-time multi-user interactions. A physiology engine governs patient dynamics and event triggers, while an interaction module interprets controller inputs to coordinate team actions. All verbal exchanges and in-simulation interactions are recorded and time-stamped.

Captured data are processed by the feedback engineer, which employs an LLM guided by structured prompts derived **Non-Technical Skills for Surgeons (NOTSS)** framework [15]. The LLM analyzes communication transcripts, identifies key behavioral patterns, and generates directed interaction graphs that quantify team structure and information flow.

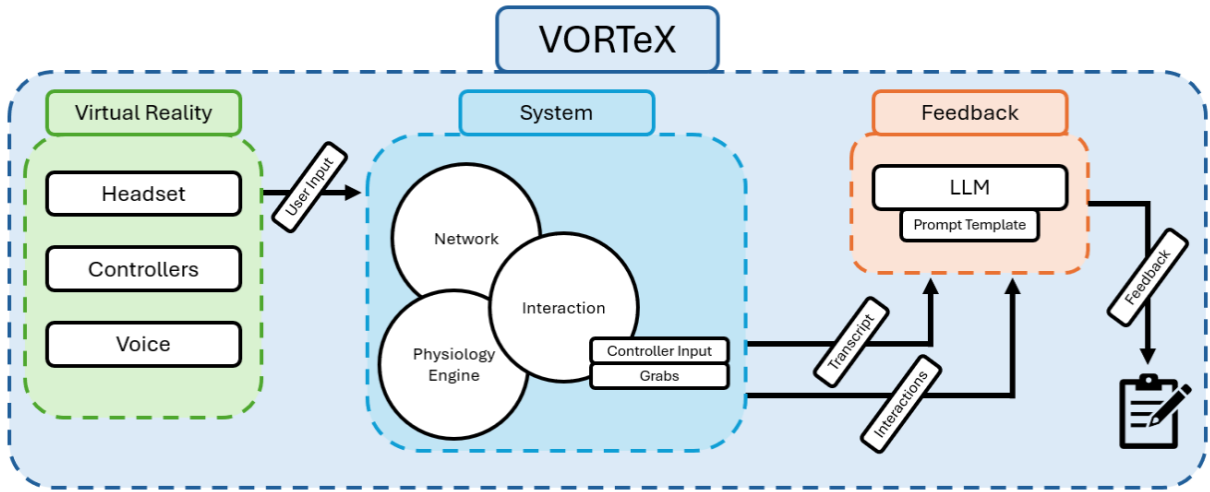

**Figure 1: VORTeX Framework diagram**

Unlike traditional simulators that rely on co-located participation and costly facilitators, VORTeX emphasizes accessibility and ease of deployment. By shifting focus from procedural fidelity to behavioral coordination, it leverages affordable consumer VR hardware to provide an immersive, scalable platform for data-driven NTS training. Figure 2 illustrates the virtual operating room environment used in the simulations, featuring a fully equipped laparoscopic setup with adjustable lighting, surgical displays, and interactable instruments. This environment reproduces the spatial layout, visual cues, and workflow of a real operating suite, promoting psychological fidelity and natural team communication while allowing distributed participants to engage synchronously from any networked location.

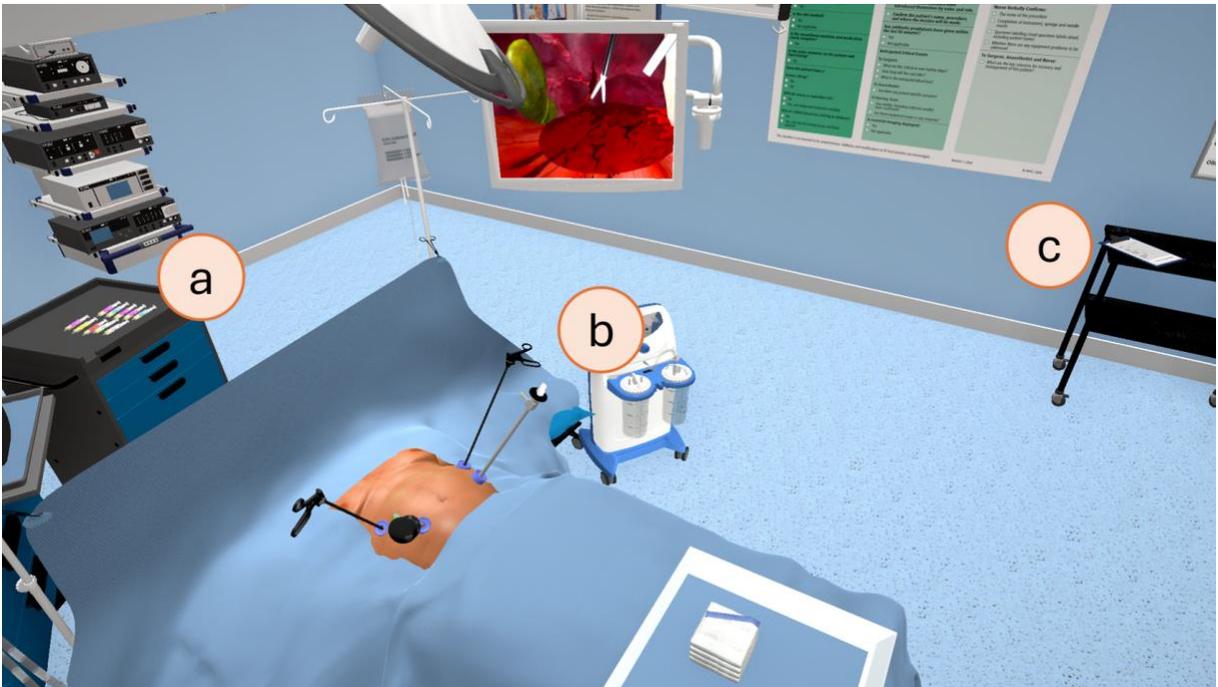

**Figure 2: The virtual operating room with a patient during a laparoscopic cholecystectomy. a) Anesthesiologist's drug cart. b) Surgeon's station. Sky arm monitors are adjustable and laparoscopic tools are interactable. c) Nurse's station showing clipboard.**

### 2.2. Simulation Design

The VORTeX simulation environment was designed to recreate the spatial layout, visual context, and cognitive demands of a real operating room while emphasizing the behavioral and communication aspects of team performance. The design goal was to elicit authentic team interactions under conditions of time pressure, uncertainty, and interdependence—key elements of non-technical skill expression. Each scenario was developed in consultation with surgeons, anesthesiologists, and perioperative nurses to ensure clinical realism, appropriate task sequencing, and accurate role responsibilities.

Two emergency scenarios were embedded within an LC procedure, reflecting the priorities of the ACS/APDS Phase III National Skills Curriculum. The **pneumothorax scenario**, as seen in Figure 3, manifests as sudden hypoxia requiring rapid cross-disciplinary recognition and communication to diagnose and manage. The **intra-abdominal bleeding scenario**, as seen in Figure 4, presents as progressive hemorrhage with escalating vital sign deterioration, eventually requiring the team to decide on conversion to open surgery. Both scenarios were developed with clinical experts to ensure realistic progression: Bleeding evolves through discrete severity stages revealed by laparoscopic exploration, while pneumothorax demands immediate coordination to stabilize ventilation. These scenarios introduce contrasting temporal pressures which elicit distinct patterns of decision making and communication under stress.

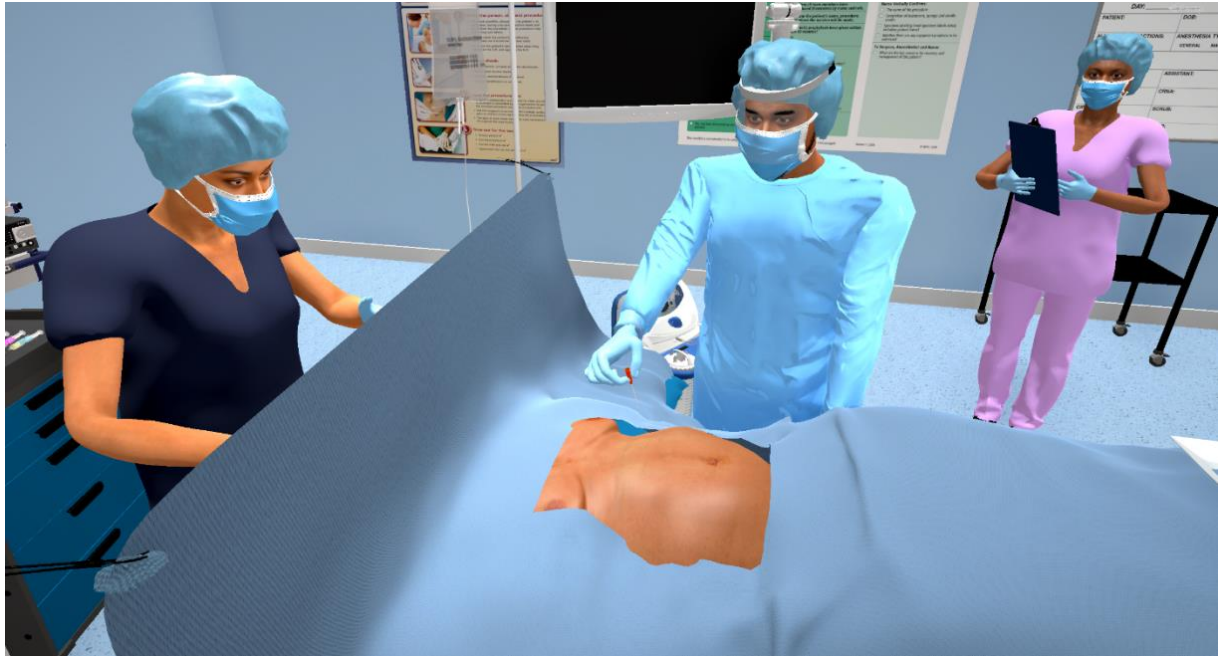

**Figure 3: Virtual operating room during pneumothorax emergency scenario.**

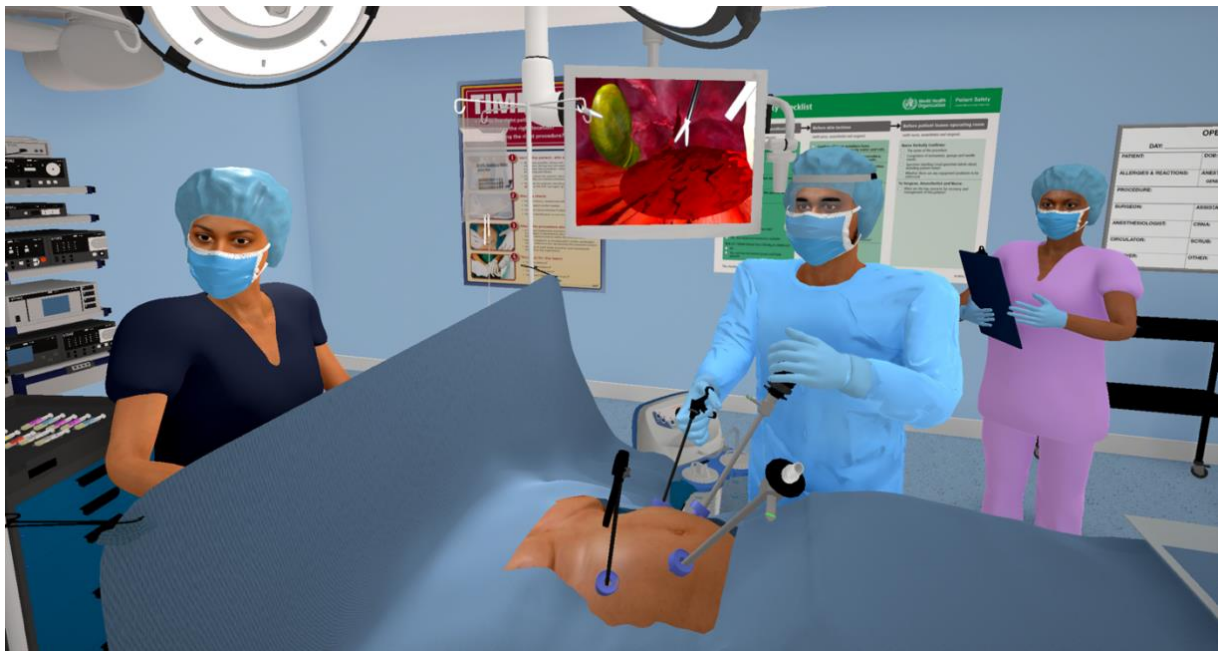

**Figure 4: Virtual operating room during intra-abdominal bleeding emergency scenario.**

The virtual operating room replicates the real-life counterpart with adjustable lighting, movable monitors, and spatial arrangements promoting natural team communication

(Figure 2). Stations within the virtual operating room have role-specific interfaces and interactables that are also designed with these principles of realism and immersion. The anesthesiologist administers pharmacologically accurate medications through an IV line that dynamically alters patient vitals (Figure 2-marked as "a"), the surgeon manipulates laparoscopic instruments via simplified tool interactions (Figure 2-marked as "b"), and the nurse controls simulation flow via a clipboard interface, initiating the timeout, requesting equipment (blood, argon, major kit), and calling for help (Figure 2-marked as "c"). All roles wear appropriate PPE, and most environmental objects are interactable to preserve immersion and provide contextual affordances for non-verbal communication.

Patient vital signs are driven by a physiology engine augmented to respond to trauma and drug onsets, ensuring that hemodynamic and respiratory responses authentically reflect clinical actions and temporal progression. Numerical values and waveform displays update in real-time based on drug administration, blood loss, and procedural interventions, creating realistic triggers for team communication. Event sequencing (e.g., bleeding severity escalation, pneumothorax onset) is pre-scripted but responsive to team performance, prompting adaptive decision-making without breaking clinical plausibility or immersion.

Performance optimization prioritized stable network synchronization and visual continuity across participants that may be geographically distributed. This was achieved through asset optimization until the 24-fps threshold for VR continuity was exceeded. Adopting a hybrid TCP/UDP network stack ensured critical state data such as patient vitals and simulation events were reliably synchronized without overburdening the network. Interaction design focused on high-frequency tracking of head movements, hand positions, and gaze vectors to support gestural communication (pointing, nodding) without requiring costly facial or finger tracking; this deliberate limitation preserved the portability of consumer-grade hardware while making most objects interactable, as participants consistently attempted environmental manipulation upon entry and non-responsiveness broke immersion.

### 2.3. Network Architecture

The multi-user aspect of VORTeX is implemented using a client-server architecture (Figure 5) that supports distributed collaboration while maintaining secure, low-latency communication. The server functions as an independent coordination node responsible for synchronizing simulation states and managing bidirectional data exchange among connected clients. To emulate a local-area network and ensure data security, a virtual private network (VPN) was used to create encrypted tunnels between clients and the central server. This configuration enables geographically dispersed users to participate in shared training sessions while preserving data integrity and minimizing latency.

The separation of client and server processes provides greater deployment flexibility, enabling the server to run on a local host or a remote machine, depending on resource availability. Separation of client and server also allows each side of the application to focus on data collection relevant to its side of the network. Ensuring low network latency is crucial for maintaining the realism of team communications in multi-user environments. To achieve this, VORTeX employs a hybrid transport strategy, combining the Transmission Control Protocol (TCP) for reliable synchronization of critical information (e.g., simulation state, patient vitals, and avatar identifiers) with the User Datagram Protocol (UDP) for high-frequency updates of avatar movements and gestures. This design prioritizes essential state synchronization while allowing lightweight, low-latency streaming of motion data, thereby ensuring smooth performance under variable network conditions.

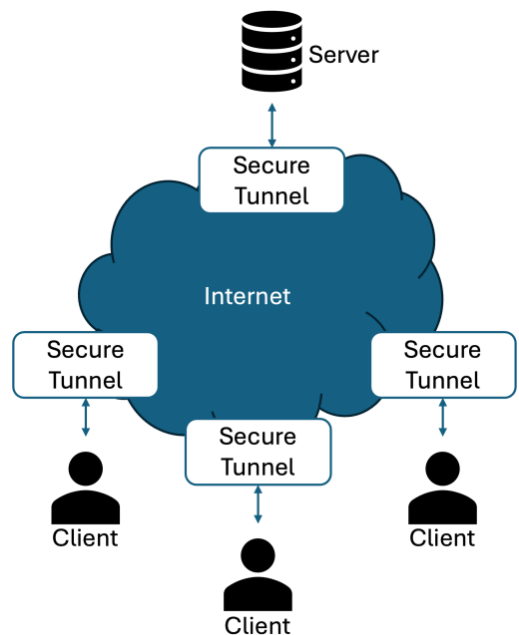

**Figure 5: Client–server communication topology illustrating encrypted VPN-based data exchange among distributed participants. Each node connects to a central coordination server via IPSec-secured tunnels, enabling low-latency, synchronized interaction across geographically dispersed sites.**

**2.4. Large Language Model**

During each VR training session, client systems continuously record multiple digital streams: Participant interactions with virtual objects, gaze vectors, and audio transcriptions of spoken dialogue. At the end of a session, these data streams are transmitted to a central server, where they are time-synchronized, anonymized and merged into a unified transcript annotated with participant identifiers and event timestamps.

To evaluate team performance within the simulated surgical environment, this consolidated transcript is analyzed by an LLM. Behavioral performance is operationalized in terms of behavioral indicators defined by the NOTSS framework [15], which categorizes observable behaviors into four domains: situational awareness, decision making, communication and teamwork, and leadership. The objective of the analysis is to detect instances of these non-technical skills, attribute them to the relevant participants, and generate a directed interaction graph representing communication flow and team dynamics.

Due to privacy and data governance concerns, the analysis was performed using a locally hosted, open-source reasoning model rather than a cloud-based API. For this work, DeepSeek-R1-0528 [16] was chosen due to its strong performance on reasoning and human-preference benchmarks and its support for detailed control over prompt structure and inference parameters. This reasoning model performs relatively well compared to other open-source and closed-source models on a number of human preference and reasoning benchmarks.

The model receives as input a composite prompt consisting of three components: (i) a prefix, which demarcates the beginning of the transcript; (ii) the transcript itself; and (iii) a postfix containing task instructions and the NOTSS behavioral rubric. The postfix instructs the model to (1) identify demonstrations of NOTSS-aligned behaviors by any participant, (2) self-verify and refine the extracted results, and (3) construct a directional interaction network in JSON format. In this network, each node corresponds to a participant, and each edge encodes an observed interaction, where the source node denotes the individual demonstrating a NOTSS skill, the target node represents the recipient of that behavior, and the edge attributes include the skill category, a timestamp, and a brief descriptive rationale. Full prompt wording and the JSON output schema are provided in Table S1 of Appendix A to ensure methodological transparency and reproducibility.

Inference parameters were configured to match those used in the model's public evaluation: temperature = 0.6, top-p = 0.95, and no additional sampling mechanisms. This configuration balances determinism and diversity in generation while maintaining reproducibility. The resulting JSON output provides a machine-readable representation of inter-participant behavioral dynamics that can be visualized or quantitatively analyzed to assess team performance.

We implement three safeguards to reduce model error and improve auditability: (i) prompt-regularized outputs (strict JSON schema with required fields), (ii) temporal consistency checks (edge timestamps must map to existing transcript spans), and (iii) internal coherence checks (role labels must match session metadata). For validity, LLM outputs are compared against expert NOTSS annotations on a subset of sessions to estimate agreement (percent agreement and Cohen's κ), and sensitivity analyses are performed over ASR confidence and prompt variants.

## 2.5. User Design Study

An Institutional Review Board–approved study was conducted during the 2024 Society of American Gastrointestinal and Endoscopic Surgeons (SAGES) conference to evaluate the usability and perceived educational value of the VORTeX simulator. Each participant completed a short pre-questionnaire to collect demographic and experiential data, followed by a ten-minute simulation session and a post-questionnaire assessing system usability and non-technical skills (NTS) perception.

Twelve surgical professionals participated (eight male, four female; ages 20–59 years). Nine were surgical residents (six PGY-1, two PGY-3, one PGY-4) and three were attending surgeons. Five participants had prior experience with team-based medical simulation training, and eight had previous exposure to virtual-reality systems.

Each participant operated as the surgeon within the simulation. Two additional medical professionals, experienced users of the system, remotely played the nurse and anesthesiologist roles from Boston, MA, while the simulation server was hosted in Cleveland, OH. This configuration demonstrated the system's ability to support geographically distributed multi-user participation.

The post-questionnaire (Tables S1 and S2 in Appendix B) consisted of 13 items. The first ten items followed the System Usability Scale (SUS) [17], assessing ease of use, learnability, and integration. The final three items measured perceived improvement in teamwork, communication, and decision-making based on the NOTSS framework [15]. Participants rated all items on a five-point Likert scale (1 = strongly disagree to 5 = strongly agree).

For interpretability, mean scores above 2.5 were considered directionally positive. Although SUS yields a single composite score, individual item responses were analyzed descriptively to explore usability trends and their correlation with self-reported confidence in non-technical skills (Q12). Correlation analyses referenced specific question numbers (Q#) to maintain reproducibility.

## 2.6 Statistical Analysis

Descriptive statistics were calculated for all questionnaire items to summarize usability and perceived NTS outcomes. Item responses were reported as means ± standard deviations on a five-point Likert scale. For exploratory analysis, Pearson correlation coefficients (r) were computed to examine associations among key variables, including prior clinical experience, age, system usability (SUS items), and self-reported confidence in NTS (Q12). Correlation strength and direction were interpreted in the context of the small sample size (n = 12), and results were considered indicative rather than inferential. Statistical computations were performed in Python 3.11 using the *pandas* and *scipy.stats* libraries.

## 3   RESULTS

The results address the two central objectives of this study: evaluating the usability, feasibility, and perceived educational value of the VORTeX multi-user virtual reality simulator, and examining the validity of the LLM–based analytics pipeline for objectively characterizing team communication and leadership behaviors during simulated operative crises. All twelve participants completed both the simulation and questionnaires with no missing data. Findings related to the usability and educational value of the simulator are presented first, including analyses of post-questionnaire responses, relationships between participant characteristics and performance metrics, and system performance outcomes (Sections 3.1–3.3). The subsequent section (3.4) focuses on the LLM-derived communication networks, providing evidence of the model's ability to capture authentic team dynamics and leadership patterns during the simulations.

### 3.1 User Experience and Perceived Educational Value

The post-questionnaire results suggest that participants generally found the simulator easy to use and perceived value in it, particularly in improving non-technical skills. The highest-rated item (Q11: $\bar{x}$ = 4.42, σ = 0.67) reflected the belief that continued use of the simulation would help enhance teamwork, communications, and decision-making. Participants also indicated that they did not need to learn a lot before getting started (Q10: $\bar{x}$ = 2.58, σ = 1.16) and did not find the system cumbersome (Q2: $\bar{x}$ = 2.50, σ = 0.90). However, confidence in using the system was relatively low (Q9: $\bar{x}$ = 2.67, σ = 1.56), pointing to a wide range of comfort levels among users and suggesting that the technology may be somewhat intimidating. Interestingly, only one-third of participants reported regular VR use, which could have contributed to this lower confidence. Nonetheless, this added challenge may have benefited the study's aim of simulating stressful conditions to foster communication and collaboration.

Despite some hesitancy in operating the system, participants' responses were otherwise positive. Many felt that the system was not unnecessarily complex (Q2: $\bar{x}$ = 2.08, σ = 0.90) and believed its various functions were well-integrated (Q5: $\bar{x}$ = 3.67, σ = 1.15). There was also a strong indication that most people would learn to use the system quickly (Q7: $\bar{x}$ = 3.75, σ = 0.97). Team performance during the laparoscopic emergency scenario received a high average rating of 4.08 (Q13: $\bar{x}$ = 4.08, σ = 0.90), suggesting that the system supported effective collaboration under pressure. Overall, while some variability existed in individual comfort and confidence levels, highlighted by higher standard deviations in those areas, the results demonstrate that participants found the simulator valuable, especially in cultivating non-technical skills.

The full distribution of post-questionnaire responses are shown in Figure 6. Median values above the midscale (2.5) for most items indicate generally favorable usability and educational value.

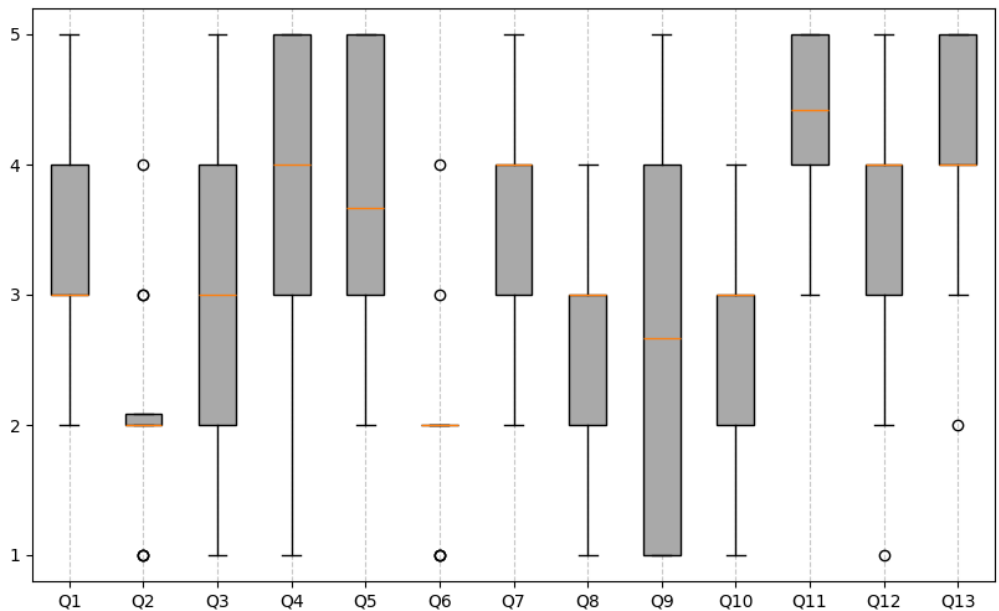

**Figure 6: Box-and-whisker plot showing response distributions for post-questionnaire items.**

**3.2 Relationships Between User Characteristics and Perceived Usability**

An exploratory Pearson correlation analysis was conducted to examine possible relationships among demographic factors, usability perceptions, and self-reported NTS confidence based on the response to the pre- and post- questionnaires. Strong inverse relationships were observed between years in clinical practice and perceived system usability: clinicians with more years of experience found the system less easy to use ($r \approx -1.0$) and reported lower confidence in using it ($r \approx -0.9$). These participants also expressed a greater perceived need for technical support ($r \approx 1.0$) and additional learning before effectively using the system ($r \approx 1.0$).

Age showed a similar but less pronounced pattern. Although older clinicians did not rate the system as markedly less usable overall, they tended to report lower confidence in their non-technical skills after using this system ($r \approx -0.7$), suggesting that reduced familiarity with immersive technologies may modestly influence perceived learning benefit.

Participants who rated the system as easy to use also tended to report greater confidence using it ($r \approx 0.8$). Confidence in using the system was positively associated with self-reported confidence in non-technical skills such as teamwork, communication, and decision-making ($r \approx 0.8$), whereas perceived system complexity negatively influenced confidence ($r \approx -0.8$). Clinicians with more years in practice rated team performance during the laparoscopic emergency lower ($r \approx -0.8$), suggesting a possible gap between system usability and perceived team effectiveness among more experienced practitioners.

Given the small sample size ($n = 12$), these relationships are reported descriptively to indicate direction and strength rather than as inferential evidence.

### 3.3 System Performance and Communication Fidelity

Consistent system performance and low communication latency are critical for maintaining immersion and the realism of team interaction in multi-user VR environments. Across all participant sessions, the simulator achieved an average frame rate of 73 fps, which maintained the 60 fps threshold for increased incidents of cybersickness [18]. The high refresh rate likely contributed to overall comfort and the favorable usability ratings (Q1: "I would like to use this system frequently," $M = 3.5$).

Equally important for preserving immersion is a low and stable latency between the server and clients. Higher latencies can result in interactions taking noticeable time to complete and more importantly, gestures and other nonverbal communication may not synchronize with spoken words. The average round-trip latency between clients and the central server was 75 ms, with a range from 20 ms during low network load to 174 ms under heavier traffic. These values fall within accepted parameters for real-time multi-user VR systems [37] and did not produce noticeable delays in motion or speech synchronization.

### 3.4 LLM-Based Analysis of Team Communication Dynamics

The LLM analysis produced valid and interpretable results across all sessions, successfully translating voice transcripts into structured communication networks. Each graph conformed to the requested schema, with nodes representing the surgeon, anesthesiologist, and nurse, and directed edges denoting initiator–receiver exchanges.

Figure 7 presents a representative LLM-generated interaction network. The directed edges capture not only who communicates with whom, but also the *function* of each interaction - situational awareness, communication and teamwork, leadership, or decision making - along with precise temporal context. In this case, the network reveals a dense, bidirectional flow of information that evolves with the clinical situation. Early in the sequence, the surgeon initiates a timeout ("Communication and Teamwork, 1:34:13"), which is immediately reinforced by the nurse's confirmation of procedure and site, reflecting shared responsibility rather than unilateral command. As the case progresses, situational awareness propagates laterally across roles: the nurse discloses a critical comorbidity ("1:35:38"), the surgeon requests and integrates status updates ("1:37:09"), and the anesthesiologist both receives and acts on this information through blood product ordering and escalation. Leadership behaviors are distributed rather than centralized, with both the surgeon and nurse independently confirming that help is en route, illustrating redundancy and resilience under stress. Overall, the structure of the network reflects anticipatory coordination, role-appropriate initiative, and timely escalation.

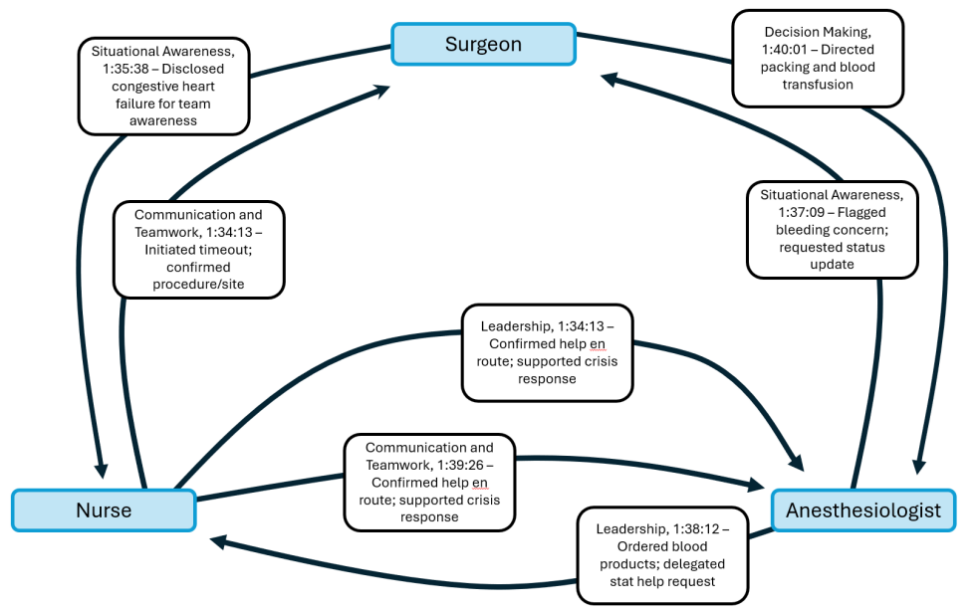

**Figure 7: Example of an LLM-generated interaction graph, illustrating directed communications among the surgeon, nurse, and anesthesiologist during a simulated intraoperative crisis. Edge attributes include behavioral category, timestamp, and rationale.**

Across all sessions, the LLM consistently captured the directionality of team interactions, distinguishing between initiating and receiving behaviors during the simulated intraoperative events. To interpret the resulting graphs, we characterized each participant's communication behavior using three standard network metrics. The **out-degree** reflects the number of distinct communication events initiated by a participant, indicating a more directive or leadership-oriented role. The **in-degree** represents the number of communications received, denoting information integration or confirmatory behavior. The **clustering coefficient** measures how interconnected a participant's communication partners were, thereby reflecting the cohesion and redundancy of local information exchange. Together, these features capture the balance between leadership, coordination, and information flow within the team.

The aggregated results are presented in Table 1. The surgeon node had the highest mean in-degree (3.84), indicating that the surgeon was the most frequent recipient of information and confirmations from other team members, consistent with the surgeon's role as the primary decision-maker in operative settings. On the other hand, the nurse node exhibited the highest mean out-degree (3.62), reflecting a tendency to initiate communication regarding patient status and operating-room coordination. The anesthesiologist had a more balanced interaction pattern (3.46 out-degree, 2.62 in-degree), serving as both an initiator and recipient of critical exchanges. The surgeon also showed the highest clustering coefficient (0.92), indicating participation in the most tightly interconnected communication subnetwork within the team.

Table 1: Mean ± standard deviation of in-degree, out-degree, and clustering coefficients for each team role in the LLM-generated interaction graphs.

|  | In-Degree (Mean, SD) | Out-Degree (Mean, SD) | Clustering Coefficient (Mean, SD) |
| --- | --- | --- | --- |
| **Surgeon** | 3.84 ±1.77 | 2.23 ±1.30 | 0.92 ±0.19 |
| **Anesthesiologist** | 2.61 ±1.45 | 3.46 ±1.13 | 0.85 ±0.23 |
| **Nurse** | 2.77 ±1.30 | 3.62 ±1.56 | 0.88 ±0.24 |

## 4  DISCUSSIONS

This work presents an integrated framework that combines immersive VR with LLM analytics to support training and evaluation of team-based non-technical skills (NTS) in the operating room. The results from the SAGES 2024 pilot study establish technical feasibility, educational plausibility, and analytic potential of this approach. Specifically, the findings demonstrate that (i) a multi-user, networked VR environment can elicit authentic intraoperative team interactions under simulated stress; and (ii) a locally hosted LLM can extract and structure these interactions into interpretable communication networks aligned with recognized NTS frameworks. Together, these components illustrate how immersive simulation and AI-driven analysis can converge to form a scalable, data-centric paradigm for NTS education.

VORTeX bridges two traditionally separate domains: high-fidelity simulation and automated behavioral analytics. The VR component provides experiential realism and psychological engagement, conditions known to enhance learning transfer, while the LLM introduces scalable, post-hoc interpretation of team communication. Unlike prior VR simulators focused primarily on psychomotor training or scripted role-play, VORTeX enables real-time, geographically distributed collaboration through consumer-grade hardware. Its integration with an LLM-based analytic backend transforms the rich but unstructured communication data into structured, machine-interpretable graphs that quantify leadership, coordination, and communication flow. This fusion demonstrates a credible pathway toward AI-augmented debriefing, addressing long-standing barriers in simulation-based education: limited facilitator bandwidth and the subjectivity of human observation.

Participants perceived the system as a valuable tool for developing teamwork, communication, and decision-making skills (Q11, $\bar{x}$ = 4.42). The VR environment replicated the stress, and coordination demands of real operative crises (e.g., pneumothorax, intra-abdominal bleeding) within a psychologically safe setting. Meanwhile, the LLM-derived communication graphs captured role-specific patterns consistent with recognized intraoperative hierarchies - surgeons as central information integrators, nurses as coordinators and initiators, and anesthesiologists as balanced intermediaries. This correspondence provides preliminary evidence of construct coherence, suggesting that the automated analytic pipeline can extract recognizable team behaviors without manual annotation. However, the evidence should be viewed as early and exploratory; without expert-coded benchmarks or inter-run reliability data, definitive claims of validity cannot yet be made.

Beyond the behavioral insights, the integration of VR and LLMs represents a methodological advance. The VR component supports distributed, synchronous simulation using accessible, consumer-grade hardware, while the analytic backend produces machine-readable summaries that could standardize debriefing and reduce facilitator dependence. By maintaining all processing locally through a privacy-compliant, open-source model (DeepSeek-R1-0528), the framework meets data governance requirements critical for educational and clinical settings. The resulting directed communication graphs introduce a reproducible quantitative language for team behavior-transforming qualitative debriefing into a structured, scalable process that can support longitudinal tracking and adaptive scenario design.

Several limitations warrant discussion. The small sample size (*n* = 12) and single-site, event-based recruitment limit statistical inference and generalizability, and the findings should be interpreted as feasibility and construct-coherence evidence rather than confirmatory validation. Participants were drawn from a self-selected group of surgical professionals at a national conference -individuals likely more technologically receptive and possibly more experienced in simulation than the general clinical population. The simulation time per participant was relatively short (10 minutes), constraining behavioral diversity and limiting the observation of adaptive team processes such as leadership transitions or strategy evolution over time.

Importantly, in this pilot study only the surgeon role was played by study participants, while the nurse and anesthesiologist roles were performed by experienced system users to ensure scenario stability, safety, and consistent progression of the simulated emergencies. As a result, the observed interaction networks reflect surgeon–team coupling within a controlled role scaffold rather than fully emergent dynamics of intact clinical teams. Consequently, conclusions regarding team behavior, leadership distribution, or relative team functionality should be interpreted as properties of the analytic framework and interaction structure, not as definitive measurements of real-world team effectiveness.

The analytic pipeline relies on automatic speech recognition (ASR) and speaker diarization, both of which can introduce transcription and segmentation errors. Misattribution of utterances or timing drift may affect the accuracy of edge assignments in the communication graph. Additionally, the mapping of dialogue segments to NOTSS categories remains heuristic and context-dependent; the LLM's interpretation of leadership, decision-making, or teamwork behaviors has not yet been benchmarked against expert human raters. Without such cross-validation, the current outputs represent plausible patterns rather than verified classifications.

Hardware tradeoffs also affected ecological fidelity. The absence of facial expression capture, eye-tracking, and fine-grained hand gestures limits the range of non-verbal cues available to the model, potentially underrepresenting affective and leadership behaviors. While these omissions were deliberate to preserve portability and cost-efficiency, they reduce the multimodal realism necessary for comprehensive NTS evaluation. Finally, the deterministic configuration of DeepSeek-R1-0528 ensured within-session consistency, but formal assessments of inter-run reproducibility, model generalizability, and bias across participant demographics remain necessary before this framework can be used for high-stakes evaluation.

Importantly, this pilot was not intended as a full implementation of the ACS/APDS Phase III Team-Based Skills Curriculum but as an early demonstration of how immersive, AI-driven analytics could support its objectives. Future studies should evaluate the platform within the Phase III framework, with larger, institutionally diverse cohorts, extended scenarios, and expert–LLM concordance analyses to establish criterion validity and educational impact.

Despite these constraints, the study establishes a credible foundation for expanding the integration of immersive simulation and AI-based analytics in surgical education. Future work should include multi-institutional cohorts, extended scenarios, and human–LLM concordance analyses to quantify reliability and validity. Linking network-derived features to objective performance outcomes, such as expert NTS ratings or error mitigation, will be critical to establish criterion validity. Incorporating multimodal signals (gaze, physiology, and high-fidelity gestures) could further refine behavioral inference. Ultimately, the combination of VR-based experiential learning and LLM-driven analysis offers a scalable, privacy-conscious, and data-rich framework for the objective assessment and development of non-technical skills in operative teams.

# 5    CONCLUSION

This study establishes the feasibility of integrating immersive virtual reality with large language model analytics for team-based training and assessment of non-technical skills in the operating room. The VORTeX platform elicited realistic communication and coordination behaviors within a psychologically safe, distributed simulation environment, while the LLM-based analytic pipeline transformed unstructured team dialogue into interpretable communication networks. Together, these components demonstrate a scalable and privacy-compliant framework for objective evaluation of teamwork and leadership in surgery. Future work will expand validation across diverse cohorts, incorporate richer non-verbal and contextual signals, and refine automated feedback mechanisms toward continuous, data-driven improvement of operative team performance.

**ACKNOWLEDGEMENTS**

This project was supported by grants from the National Institutes of Health (NIH)/NIBIB R01EB005807, R01EB033674, and R01EB032820.

**Appendix A:**

**Table S1: LLM prompt components.**

| | |
|---|---|
| Prefix | <\|User\|>[BEGIN TRANSCRIPT]\n |
| Transcript | |
| Postfix | \n[END TRANSCRIPT]\nThe Non-Technical Skills for Surgeons is a behavioral assessment tool for evaluation and improvement of behaviors in the operating room. The tool outlines four distinct categories of non-technical skills, each with corresponding elements that define the category.\nThese categories and their elements are as follows:\n- Situational awareness as evidenced by gathering surgery related information, understanding information, projecting and anticipating future state\n- Decision making as evidenced by considering options, selecting and communicating option, and implementing and reviewing decisions\n- Communication and teamwork as evidenced by establishing a shared understanding of and coordinating surgical team activities\n- Leadership as evidenced by setting and maintaining standards for the surgical team, coping with pressure, and supporting others\n\nYour goal is to complete the following tasks that compound upon themselves. Complete them in order and think carefully about your responses.\n1) Use the transcript of the simulated surgical environment to observe for demonstration of the non-technical skills within the surgical team as described by the Non-Technical Skills for Surgeons behavioral assessment tool. Apply the assessment tool to each of the participants, even if they are not playing the role of a surgeon.\n2) Check your response for accuracy. Make corrections if needed.\n3) Create a directional network based on the results of the assessment. Use the following rules for the creation of the directional network: The nodes of the network are the participants and the edges are the results from the assessment. The source node for each edge is the participant demonstrating the non-technical skill. The destination node for each edge is the other participant involved in the demonstration. The label for each edge is the non-technical skill being demonstrated. The description for each edge is a timestamp referencing where the demonstration occurs in the source transcript along with a short explanation. Create the network in a JSON format, suitable for importing into a network visualization tool.<\|Assistant\|> < think> |

**Appendix B:**

### Table S1: SUS questions in the post-questionnaire.

| No. | Question |
|---|---|
| Q1 | I think that I would like to use this system frequently. |
| Q2 | I found the system unnecessarily complex. |
| Q3 | I thought the system was easy to use. |
| Q4 | I think that I would need the support of a technical person to be able to use this system. |
| Q5 | I found the various functions in this system were well integrated. |
| Q6 | I thought there was too much inconsistency in this system. |
| Q7 | I would imagine that most people would learn to use this system very quickly. |
| Q8 | I found the system very cumbersome to use. |
| Q9 | I felt very confident using the system. |
| Q10 | I needed to learn a lot of things before I could get going with this system. |

### Table S2: NTS questions in the post-questionnaire.

| No. | Question |
|---|---|
| Q11 | I believe continued use of this system would help improve my non-technical (teamwork, communications, decision-making) skills. |
| Q12 | I feel more confident in my non-technical (teamwork, communications, decision-making) skills after using this system. |
| Q13 | The team performed well together in addressing the laparoscopic emergency. |